\documentstyle[12pt]{article}
\begin{document}\begin{titlepage}
\vspace*{1cm}

\begin{center}
{\Large \bf Exact Results in Renormalization of  Softly\\[0.2cm]
Broken SUSY Gauge Theories}\footnote{ Talk presented at XXIX
International Conference on High Enegy Physics (Vancouver, Canada,
July, 1998) and at XI International Conference "Problems of Quantum
Field Theory" (Dubna, Russia, July, 1998)}

\vspace*{1cm}

{\large D.I.~Kazakov}
\date{}

\vspace*{1cm}

{\it Bogoliubov Laboratory of Theoretical Physics,
Joint Institute for Nuclear Research, \\
141 980 Dubna, Moscow Region, RUSSIA, \\
E-mail: kazakovd@thsun1.jinr.ru}
\end{center}
\vspace*{2cm}

\begin{abstract}
It is shown that softly broken theory is equivalent to  a
 rigid theory in external spurion superfield. The singular part of
effective action in a broken theory follows from a rigid one by a
simple redefinition of the couplings.  This gives an explicit  relation
 between  the soft and rigid couplings renormalizations.  As an
illustration the renormalization group functions  in the MSSM have been
calculated.  The method opens  a  possibility to construct a totally
all loop finite N=1 SUSY gauge theory, including the soft SUSY breaking
terms.  Explicit relations between the soft terms, which lead to a
completely finite theory in any loop order, are given.
\end{abstract}
\end{titlepage}

\section{Introduction}

The gauge theory  with softly broken supersymmetry has been widely
studied.
A powerful method  which keeps supersymmetry
manifest is the supergraph technique~\cite{supergraph,supergraph2}. It is
also applicable to softly broken SUSY models by using the "spurion" external
superfields~\cite{spurion,spurion2,Scholl}. As has been shown by
Yamada~\cite{Yamada} with the help of the spurion method the calculation of
the $\beta$ functions of soft SUSY-breaking terms is a much simpler task
than in the component approach.

In a recent paper~\cite{AKK} we have developed  a modification of the spurion technique
in gauge theories
 and have  formulated the Feynman rules. We have shown that the
ultraviolet divergent parts of the Green functions of a softly broken
SUSY gauge theory are proportional to those  of a rigid theory with the
spurion fields factorized.

The main idea is that a softly broken supersymmetric gauge theory
can be considered
as a rigid SUSY theory imbedded into  external space-time independent
superfield, so that all  couplings and masses
become external superfields.

 \underline{The main Statement:}

\begin{center}
 {\bf Softly Broken  SUSY Theory \ \ $\approx$ \ \  Rigid SUSY Theory
in External  Field}

{\bf  The Coupling  $g$ \ \  $\Rightarrow $ \hspace{0.2cm}
External Superfield $\Phi_0$ \phantom{\ \  }}
\end{center}

\vspace{0.1cm}

 \underline{Consequence:}
 Singular part of effective action depends on external superfield, but
not on its derivatives:
\nopagebreak[4]
\begin{picture}(20,20)(160,50)
\put(240,20){\line(1,1){20}}
\put(240,40){\line(1,-1){20}}
\put(270,20){\line(1,1){20}}
\put(270,40){\line(1,-1){20}}
\put(310,20){\line(1,1){20}}
\put(310,40){\line(1,-1){20}}
\end{picture}
$$S_{eff}^{sing}(g) \ \ \Rightarrow \ \ S_{eff}^{sing}(\Phi_0,
 D^2\Phi_0, \bar D^2\Phi_0, D^2\bar D^2\Phi_0)$$

This approach to a softly broken sypersymmetric theory allows us to use
remarkable mathematical properties of $N=1$ SUSY theories such as
non-renormalization theorems, cancellation of quadratic divergences, etc.
We show that the renormalization procedure in a softly broken SUSY
gauge theory can be performed in exactly the same way as in a rigid theory
with the renormalization constants being external superfields. They are
related to the corresponding  renormalization constants of a rigid theory by
the coupling constants redefinition. This  allows us to find explicit
relations between the  renormalizations  of soft  and rigid  couplings.

 Throughout the paper we assume the existence of some gauge and SUSY
invariant regularization and a minimal subtraction procedure.

As an application of the above mentioned relations we consider a
possibility of constructing totally finite supersymmetric theories
including the soft breaking terms. This problem has already been
discussed several times~\cite{Mezinchescu,Jones}.   We  show
that by
choosing the soft terms in a proper way one can reach complete all loop
finiteness.  Moreover, there is no new fine-tuning.  The soft terms are
fine-tuned in exactly the same way as the corresponding Yukawa
couplings~\cite{Finlet}.

\section{Softly Broken $N=1$ SUSY Gauge Theory }

 Consider  pure $N=1$ SUSY Yang-Mills theory with a simple
gauge group. The Lagrangian  of a rigid theory is given by
\begin{eqnarray}\hspace*{-0.4cm}
 {\cal L}_{rigid} &=& \int d^2\theta~\frac{1}{4g^2}{\rm
Tr}W^{\alpha}W_{\alpha} + \int d^2\bar{\theta}~\frac{1}{4g^2}{\rm
Tr}\bar{W}^{\alpha}\bar{W}_{\alpha} ~.
\label{rigidlag}
\end{eqnarray}

To  perform a soft SUSY breaking, one can introduce a gaugino mass
term
\begin{equation}
 -{\cal L}_{soft-breaking} =
\frac{m_A}{2}\lambda\lambda + \frac{m_A}{2}\bar \lambda \bar \lambda,
\end{equation}
where $\lambda$ is  the gaugino field. To rewrite it in terms of
superfields, let us  introduce an external spurion superfield
 $\eta=\theta^2$, where $\theta$ is a Grassmannian parameter. The
softly broken Lagrangian  can now be written as
\begin{eqnarray}
{\cal L}_{soft} &=& \int d^2\theta~\frac{1}{4g^2}(1-2\mu\theta^2) {\rm
Tr}W^{\alpha}W_{\alpha}  + \int
 d^2\bar{\theta}~\frac{1}{4g^2}(1-2\bar{\mu}\bar{\theta}^2) {\rm
Tr}\bar{W}^{\alpha}\bar{W}_{\alpha}.  \label{soft}
\end{eqnarray}
In terms of component fields the interaction with external spurion
 superfield leads to a gaugino mass equal to $m_A=\mu$, while the gauge
field remains massless.  This external chiral superfield can be
considered as a vacuum expectation value of a dilaton superfield
emerging from supergravity, however, this is not relevant to further
consideration.

 As has been shown in~\cite{AKK} the Feynman rules corresponding to the
Lagrangian (\ref{soft}) needed for the calculation of the singular part
of effective action are the same as in a rigid theory with  the
substitution:
\begin{eqnarray}
g^2 \ \to \ \tilde{g}^2 = g^2\left(1 +
\mu\theta^2 + \mu\bar{\theta}^2 + 2\mu^2\theta^2 \bar{\theta}^2
\right).  \label{Tildeg}
\end{eqnarray}
Then the vector propagator in a softly broken theory is:
\begin{eqnarray}
\left<V(x_1,\theta_1,\bar{\theta_1})V(x_2,\theta_2,\bar{\theta_2})
\right>_{soft}=  \frac{{\tilde g}^2}{g^2} \left<V(x_1,\theta_1,
\bar{\theta_1})V(x_2,\theta_2,\bar{\theta_2})\right>_{rigid}
+ {\rm irrel.terms}, \label{Pvs}
\end{eqnarray}
where by irrelevant terms we mean the ones decreasing faster than
$1/p^2$ for large $p^2$.
The same is true for the ghost fields
\begin{equation}
\left<G(z_1)\bar{G}(z_2)\right>_{soft} =
\left(\tilde{g}^2/g^2\right) \left<G(z_1)\bar{G}(z_2)\right>_{rigid}
 +{\rm irrel.~terms}, \label{Gps}
\end{equation}
where $G$ stands for any ghost superfield.

Hence, to perform the analysis of the divergent part of the diagrams in
a soft theory, one has to use the same propagators as in a rigid theory
multiplied by the factor ${\tilde g}^2/g^2$. It is also obviously true
for any vertex of the ghost-vector interactions of the softly broken
theory. Each vertex of this type has to be
multiplied by the inverse factor $g^2/{\tilde g}^2$.
 The situation is less obvious with the vector vertices, however it
happens to be true in this case as well.

Thus, we see that any element of the Feynman rules for a softly broken
theory  coincides with the corresponding element of a rigid theory
multiplied by a common factor which is a polynomial in the grassmann
coordinates.

Consider now a rigid SUSY gauge theory with chiral matter.  The Lagrangian
written in terms of superfields looks like
\begin{equation}
{\cal L}_{rigid}  = \int d^2\theta d^2\bar{\theta} ~\bar{\Phi}^i
(e^{V})^j_i\Phi_j + \int d^2\theta ~{\cal W} + \int d^2\bar{\theta}
~\bar{\cal W},
\end{equation}
where the superpotential ${\cal W}$ in a  general form is
\begin{equation} {\cal  W}=\frac{1}{6}\lambda^{ijk}\Phi_i\Phi_j\Phi_k
+\frac{1}{2} M^{ij}\Phi_i\Phi_j.\label{rigid}
\end{equation}
 The SUSY  breaking terms which satisfy the requirement of "softness"
can be written as
\begin{eqnarray}
-{\cal L}_{soft-breaking}&=&\left[\frac 16 A^{ijk} \phi_i\phi_j\phi_k+
\frac 12 B^{ij}\phi_i\phi_j +h.c.\right]
+(m^2)^i_j\phi^{*}_i\phi^j,\label{sofl}
\end{eqnarray}

Like in the case of a pure gauge theory the soft terms (\ref{sofl}) can be
written down in terms of superfields by using the external spurion
field.  The full  Lagrangian for the softly broken theory can be
written as
\begin{eqnarray}
{\cal L}_{soft}&=&\int d^2\theta
d^2\bar{\theta} ~~\bar{\Phi}^i(\delta^k_i -(m^2)^k_i\eta
\bar{\eta})(e^V)^j_k\Phi_j   \label{ssofl2} \\ &+& \int  d^2\theta
\left[\frac 16 (\lambda^{ijk}-A^{ijk} \eta)\Phi_i\Phi_j\Phi_k
+ \frac 12
(M^{ij}-B^{ij}\eta ) \Phi_i\Phi_j \right] +h.c.  \nonumber
\end{eqnarray}

The Lagrangian (\ref{ssofl2}) allows one to write down the
Feynman rules for the matter field propagators and vertices in a soft
theory.
\begin{equation} \hspace*{-0.1cm}
\left<\Phi(z_1)_i\bar{\Phi}(z_2)^j\right>_{soft}=
 \label{prphi}(\delta^k_i +\frac 12(m^2)^k_i\eta \bar{\eta})
\left<\Phi(z_1)_k\bar{\Phi}(z_2)^l\right>_{rigid}
(\delta^j_l +\frac 12(m^2)^j_l\eta \bar{\eta})
 +{\rm i.t.},
\end{equation}
The vector-matter vertices,  according to eq.(\ref{ssofl2}), gain
the factor $(\delta^j_i -(m^2)^j_i\eta \bar{\eta})$ so that if in a
diagram one has an equal number of chiral propagators and
vector-matter vertices the spurion factors cancel.

The chiral vertices of a soft theory, as it follows from
eq.(\ref{ssofl2}), are the same as in a rigid theory with the
Yukawa couplings being replaced by
$$
\lambda^{ijk} \to \lambda^{ijk}-A^{ijk}\eta, \ \ \ \bar \lambda_{ijk}
\to
 \bar \lambda_{ijk} - \bar A_{ijk} \bar{\eta}.
$$
The structure of the UV counterterms in chiral vertices is similar to
that of the vector vertices, but is simpler due to the absence of
the covariant derivatives on external lines.
 Effectively the corrections to the propagator (\ref{prphi}) may be
associated with the chiral vertices, which allows one to reduce all
soft term corrections to the modification of the couplings.

\section{Renormalization of Soft versus Rigid Theory:
\protect \\ the General Case}

 The external field construction described above allows one to write
down the renormalization of soft terms starting from the known
renormalization of a rigid theory without any new diagram calculation.
The following statement is valid~\cite{AKK}:  

\newtheorem{statement}{Statement}
\begin{statement}
{\it Let a rigid theory
(\ref{rigidlag},\ref{rigid}) be renormalized via introduction of the
renormalization constants $Z_i$, defined within some minimal subtraction
massless scheme. Then, a softly broken theory (\ref{soft},\ref{ssofl2})
is renormalized via introduction of the renormalization superfields
$\tilde{Z}_i$ which are related to $Z_i$ by the coupling constants
redefinition
\begin{equation} \tilde{Z}_i(g^2,\lambda ,\bar \lambda)=
Z_i(\tilde{g}^2,\tilde{\lambda},\tilde{\bar \lambda}),
\label{Z} \end{equation} where the redefined couplings are
($\eta=\theta^2, \ \bar{\eta}=\bar{\theta}^2$)
\begin{eqnarray}
\tilde{g}^2&=&g^2(1+\mu \eta+\bar \mu \bar{\eta}+2\mu\bar \mu \eta
\bar{\eta}),
\label{g}\\
\tilde{\lambda }^{ijk}&=&
\lambda^{ijk}-A^{ijk}\eta
+\frac 12
(\lambda^{njk}(m^2)^i_n +\lambda^{ink}(m^2)^j_n+\lambda^{ijn}(m^2)^k_n)\eta
\bar \eta, \label{y1}   \\
\tilde{\bar \lambda }_{ijk}&=&
\bar \lambda_{ijk} - \bar A_{ijk} \bar{\eta}
+ \frac 12
(\bar \lambda_{njk}(m^2)_i^n
+\bar \lambda_{ink}(m^2)_j^n+\bar \lambda_{ijn}(m^2)_k^n)\eta \bar\eta
\label{y2}
\end{eqnarray} }
\end{statement}

From eqs.(\ref{Z}) and (\ref{g}-\ref{y2}) it is possible to write down
an explicit differential operator which has to be applied to the $\beta $
functions of a rigid theory in order to get those for the soft terms.

Consider first the gauge couplings $\alpha_i$. One has
\begin{equation}
\alpha_{i}^{Bare}=Z_{\alpha i}\alpha_i \ \ \Rightarrow \ \
\tilde{\alpha}_{i}^{Bare}=\tilde{Z}_{\alpha i}
\tilde{\alpha}_i,  \label{alpha}
\end{equation}
where $Z_{\alpha i}$ is the product of the wave function and vertex
renormalization constants.

Though $\tilde{\alpha}_i$ and $\tilde{Z}_{\alpha i}$ are general superfields, one has to
consider only their chiral or antichiral parts.
 The chiral part of eq.(\ref{alpha}) is
$$
\alpha_{i}^{Bare}(1+m_{A_i}^{Bare}\eta)=\alpha_i(1+m_{A_i}\eta)Z_{\alpha
i}(\tilde{\alpha})\vert_{\bar \eta=0}.
$$
Expanding over $\eta$  one has
\begin{eqnarray}
\alpha_{i}^{Bare}&=&\alpha_i Z_{\alpha i}(\alpha), \label{ab} \\
m_{A_i}^{Bare}\alpha_{i}^{Bare}&=&m_{A_i}\alpha_i Z_{\alpha i}(\alpha
)+\alpha_i D_1 Z_{\alpha i}, \label{m}
\end{eqnarray}
where the operator  $D_1$ extracts the linear w.r.t. $\eta$ part of
$Z_{\alpha i} (\tilde{\alpha})$. Due to eqs.(\ref{g}-\ref{y2}) the explicit
form of $D_1$ is
$$
D_1=m_{A_i}\alpha_i\frac{\partial}{\partial \alpha_i} \ .  \label{D1}
$$
Combining eqs.(\ref{ab}) and  (\ref{m}) one gets
\begin{equation}
m_{A_i}^{Bare}=m_{A_i}+ D_1 \ln Z_{\alpha i} \ \ \ .  \label{mgaug}
\end{equation}

To find the corresponding $\beta$ functions one has to differentiate
eqs.(\ref{ab}) and (\ref{mgaug}) w.r.t. the scale factor having in mind that
the operator $D_1$ is scale invariant. This gives
\begin{equation}
\beta_{\alpha i}=\alpha_i \gamma_{\alpha i}, \ \ \
 \beta_{m_{A i}}=D_1\gamma_{\alpha i} ,\label{ai}
\end{equation}
where $\gamma_{\alpha i}$ is the logarithmic derivative of $\ln Z_{\alpha
i}$ equal to the anomalous dimension of the vector superfield in some
particular gauges.

One can make also the transition from a rigid to a broken theory at the level of the renormalization
group equation. Namely take
\begin{equation}
\dot{\alpha}= \beta_\alpha(\alpha) \ \
\Rightarrow \ \ \dot{\tilde{\alpha}}= \beta_{\tilde{\alpha}}
(\tilde{\alpha}), \label{alp}
\end{equation}
 and expand over $\theta^2$. Then one immediately reproduces
eq.(\ref{ai}) for $m_A$.

One can go even further and consider a solution of the RGE. Then one
has in a rigid theory
\begin{equation}
\int^\alpha \frac{d\alpha'}{\beta(\alpha')} =
\log\left(\frac{Q^2}{\Lambda^2}\right).
\end{equation}
Making a substitution $\alpha \to \tilde{\alpha}$ one has
\begin{equation}
\int^{\tilde{\alpha}} \frac{d\alpha'}{\beta(\alpha')} =
\log\left(\frac{Q^2}{\tilde{\Lambda}^2}\right),
\end{equation}
where $\tilde{\Lambda}=\Lambda(1+c\theta^2+...)$. Expanding over
$\theta^2$ one finds
\begin{equation}
\frac{m_A\alpha}{\beta(\alpha)}=const .
\end{equation}
This result is in complete correspondence with
that of Ref.~\cite{Shifman}.

The same procedure can be applied for the other soft terms. This needs
the modification of the operator $D_1$ to include the $A$ parameter
 from the chiral vertex. As for the mass square terms, they need the
second order differential operator.  Relations between the rigid and
soft terms renormalizations in general case are summarized below.
Similar results on the soft terms renormalization have been obtained in
ref.\cite{NewJ} \vspace*{-0.3cm}\begin{table}[htb]
\begin{center}
\begin{tabular}{|l|l|}
\hline \hline  & \\[-0.2cm]
\hspace*{1.5cm}  The Rigid Terms & \hspace*{1.5cm}  The Soft Terms \\[0.2cm]
\hline  & \\
$\beta_{\alpha_i} =  \alpha_i\gamma_{\alpha_i}$ &
$\beta_{m_{A i}}=D_1\gamma_{\alpha i}$ \\[0.3cm]
$\beta_{M}^{ij} =\frac{1}{2}(M^{il}\gamma^j_l+M^{lj}\gamma^i_l) $&
$\beta_{B}^{ij} = \frac{1}{2}(B^{il}\gamma^j_l+B^{lj}\gamma^i_l)-
(M^{il}D_1\gamma^j_l+M^{lj}D_1\gamma^i_l) $\\[0.3cm]
$\beta_{y}^{ijk} =\frac{1}{2}(y^{ijl}\gamma^k_l+y^{ilk}\gamma^j_l+
y^{ljk}\gamma^i_l)$ &
$\beta_{A}^{ijk} = \frac{1}{2}(A^{ijl}\gamma^k_l+A^{ilk}\gamma^j_l+
A^{ljk}\gamma^i_l)$ \\[0.2cm] &
\hspace*{1cm}
$-(y^{ijl}D_1\gamma^k_l+y^{ilk}D_1\gamma^j_l+y^{ljk}D_1\gamma^i_l)$\\[0.2cm]
&  $(\beta_{m^2})^i_j=D_2\gamma^i_j $ \\[0.3cm]
\hline
\multicolumn{2}{|l|}  {} \\[-0.2cm]
\multicolumn{2}{|l|}
{$D_1= m_{A_i}\alpha_i\frac{\displaystyle \partial}{\displaystyle \partial
\alpha_i} -A^{ijk}\frac{\displaystyle \partial}{\displaystyle \partial
y^{ijk}}\ \ , \hspace{1.7cm}
\bar{D}_1=m_{A_i}\alpha_i\frac{\displaystyle  \partial}{\displaystyle
\partial \alpha_i} -A_{ijk}\frac{\displaystyle \partial}{\displaystyle
\partial y_{ijk}} $ }\\[0.3cm]
\multicolumn{2}{|l|} {$D_2= \bar{D}_1 D_1 +
m_{A_i}^2\alpha_i\frac{\displaystyle \partial }{\displaystyle \partial
\alpha_i} $} \\[0.3cm] \multicolumn{2}{|l|} {\hspace*{0.1cm}
$+\frac{1}{2}(m^2)^a_n\left(y^{nbc}\frac{\displaystyle \partial
}{\displaystyle \partial y^{abc}} +y^{bnc}\frac{\displaystyle \partial
 }{\displaystyle \partial y^{bac}}+ y^{bcn}\frac{\displaystyle \partial
}{\displaystyle \partial y^{bca}}+ y_{abc}\frac{\displaystyle \partial
}{\displaystyle \partial y_{nbc}}+ y_{bac}\frac{\displaystyle \partial
}{\displaystyle \partial y_{bnc}}+ y_{bca}\frac{\displaystyle \partial
}{\displaystyle \partial y_{bcn}}\right)$ }\\[0.3cm] \hline \hline
\end{tabular}
\end{center}
\caption{Relations between the  rigid and soft term renormalizations in a
massless minimal subtraction scheme}
\end{table}

\vspace*{-0.1cm}\section{Illustration}

To make the above formulae more clear and to demonstrate how they
work in practice, we consider the renormalization group functions in a
general theory in one loop. We follow the notation of
ref.~\cite{Jones} except that our $\beta $ functions are half of
those.  Note that all the calculations in
ref.~\cite{Jones} are performed in the framework of dimensional reduction
and the $\overline{MS}$ scheme.

The gauge $\beta $ functions and the anomalous dimensions of matter
superfields in a massless scheme are the functions of dimensionless
gauge and Yukawa couplings of a rigid theory.
In the one-loop order, the renormalization group functions of a rigid
theory are (for simplicity, we consider the case of a single gauge
coupling)\footnote{
To simplify the formulas hereafter we use the following notation:
$\alpha_i = {g^2_i}/{16\pi^2}, \ \ y^{ijk}={\lambda^{ijk}}/{4\pi}, \ \
A^{ijk}={A^{ijk}}/{4\pi}.$}
\begin{eqnarray}
\gamma_\alpha^{(1)}&=&\alpha Q, \ \ \
Q=T(R)-3C(G), \\ \gamma^{i\ (1)}_j&=& \frac{1}{2}y^{ikl}y_{jkl}-2\alpha
C(R)^i_j,
\end{eqnarray}
where $T(R),\ C(G)$ and $C(R)$ are the Casimir operators.
Using the formulae from the table we construct the renormalization group
functions for the soft terms
\begin{eqnarray}
\beta_{m_A}^{(1)} &=& \alpha m_AQ, \\
\beta_B^{ij\ (1)} &=& \frac{1}{2}B^{il}(\frac{1}{2}y^{jkm}y_{lkm}
-2\alpha C(R)^j_l)  \\
&&+ M^{il}(\frac{1}{2}A^{jkm}y_{lkm}+2\alpha m_AC(R)^j_l)
+(i\leftrightarrow j), \nonumber\\
\beta_{A}^{ijk\ (1)}&=&\frac{1}{2}A^{ijl}(\frac{1}{2}y^{kmn}y_{lmn}
-2\alpha C(R)^k_l)  \\
&&+ y^{ijl}(\frac{1}{2}A^{kmn}y_{lmn}+2\alpha m_AC(R)^k_l) \nonumber \\
&& +(i\leftrightarrow j) +(i\leftrightarrow k), \nonumber\\
\left[\beta_{m^2}\right]^{i\ (1)}_j &=&
\frac{1}{2}A^{ikl}A_{jkl}-4\alpha m_A^2C(R)^i_j  \\
&&\hspace*{-1.7cm}+\frac{1}{4}y^{nkl}(m^2)^i_ny_{jkl}
+\frac{1}{4}y^{ikl}(m^2)^n_jy_{nkl}+\frac{4}{4}y^{isl}(m^2)^k_sy_{jkl}
\nonumber.
\end{eqnarray}
One can easily see that the resulting formulae coincide with those of
ref.~\cite{Jones}. The same procedure works in higher orders of PT.

\section{Soft Renormalizations in the MSSM}

The general rules described in the previous section can be applied to any
model, in particular to the MSSM. In the case when the field content
and the Yukawa interactions are fixed, it is more useful to deal with
numerical rather than with tensor couplings. Rewriting the
superpotential (\ref{rigid}) and the soft terms (\ref{sofl}) in terms
of group invariants, one has
\begin{equation}
{\cal W}_{SUSY}=\frac{1}{6}\sum_a y_a\lambda^{ijk}_a
\Phi_i\Phi_j\Phi_k+\frac{1}{2}\sum_b M_bh^{ij}_b\Phi_i\Phi_j, \label{pot}
\end{equation}
and
\begin{equation}\hspace*{-0.1cm}
 -{\cal L}_{soft}=\left[\frac 16 \sum_a {\cal A}_a
\lambda^{ijk}_a \phi_i\phi_j\phi_k+ \frac 12 \sum_b {\cal
B}_bh^{ij}_b\phi_i\phi_j
+  \frac 12 m_{A_j}\lambda_j\lambda_j+h.c.\right]
+(m^2)^j_i\phi^{*i}\phi_j,\label{sof}
\end{equation}
where we have introduced numerical couplings $y_a,M_b,{\cal A}_a$ and
${\cal B}_b$.

Usually, it is assumed that the soft terms obey the universality
hypothesis, i.e. they repeat the structure of a superpotential, namely
\begin{equation}
{\cal A}_a=y_aA_a,\ \ {\cal B}_b=M_bB_b, \ \ (m^2)^i_j=m^2_i\delta^i_j.
\end{equation}

The  renormalization group $\beta$ functions of a rigid theory
 are (for simplicity, we assume the
diagonal renormalization of matter superfields)
\begin{eqnarray}
\beta_{\alpha_j} &=& \beta_j \equiv \alpha_j\gamma_{\alpha_j}, \\
\beta_{y_a} &=& \frac{1}{2}y_a \sum_iK_{ai}\gamma_i, \\
\beta_{M_b} &=& \frac{1}{2}M_b\sum_iT_{bi}\gamma_i,
\end{eqnarray}
where $\gamma_i$ is the  anomalous dimension of the superfield $\Phi_i$,
$\gamma_{\alpha_j}$ is the anomalous dimension of the gauge superfield
(in some gauges) and numerical matrices $K$ and $T$ specify which
particular fields contribute to a given term in eq.(\ref{pot}).

Applying  the
algorithm of the previous section
the renormalizations of the soft terms are expressed through those
of a rigid theory in the following way:
\begin{eqnarray}
\beta_{m_{A_j}}&=&D_1\gamma_{\alpha_j}, \label{ma}\\
\beta_{A_a}&=&-D_1\sum_i K_{ai}\gamma_i, \label{A}\\
\beta_{B_b}&=&-D_1\sum_i T_{bi}\gamma_i, \label{B}\\
\beta_{m^2_i}&=&D_2\gamma_i,  \label{m2}
\end{eqnarray}
and the operators $D_1$ and $D_2$ now take the form
\begin{eqnarray}
D_1&=&m_{A_i}\alpha_i \frac{\partial }{\partial \alpha_i}
 -A_aY_a\frac{\partial }{\partial Y_a},  \label{d1}\\
D_2&=&(m_{A_i}\alpha_i \frac{\partial }{\partial \alpha_i}
 -A_aY_a\frac{\partial }{\partial Y_a})^2
+m_{A_i}^2\alpha_i\frac{\partial }{\partial \alpha_i} +
m^2_iK_{ai}Y_a\frac{\partial }{\partial Y_a}. \label{d2}
\end{eqnarray}
where we have used the notation $Y_a\equiv y_a^2.$

To illustrate these rules, we consider as an example one loop
renormalization of the MSSM couplings. Leaving for simplicity the
third generation Yukawa couplings only, the superpotential is
\begin{eqnarray}
{\cal W}_{MSSM} &=&(y_tQ^jU^cH_2^i+y_bQ^j{D'}^{c}H_1^i+y_\tau L^jE^cH_1^i
+\mu H^i_1H^j_2)\epsilon_{ij},
\end{eqnarray}
where $Q,U,D',L$ and $E$ are quark doublet, up-quark, down-quark, lepton
doublet and lepton singlet superfields, respectively, and $H_1$ and
$H_2$ are Higgs doublet superfields. $i$ and $j$ are the $SU(2)$
indices.

The soft terms have a universal form
\begin{eqnarray}
-{\cal L}_{soft-breaking}&=&
\sum_im^2_i|\phi_i|^2+(\frac{1}{2}\sum_a\lambda_a\lambda_a  \\
&+& A_ty_tq^ju^ch_2^i
  +A_by_bq^j{d'}^{c}h_1^i+A_\tau y_\tau l^je^ch_1^i+
B\mu h^i_1h^j_2 ) + h.c.,  \nonumber
\end{eqnarray}
where the small letters denote the scalar components of the
corresponding superfields and $\lambda_a$ are the gauginos. The $SU(2)$
indices are suppressed.

 For illustration we calculate here  the one loop soft term
renormalizations and the gaugino mass renormalization  at the three
loop level out of a corresponding rigid $\beta $ functions.

Renormalizations in a rigid theory in the one loop order are given by
the formulae
\begin{eqnarray} \gamma_{\alpha_i}^{(1)}&=&b_i \alpha_i,
\ \ \ i=1,2,3, \ \ \ b_i={\frac{33}{5}, 1,-3}, \\
\gamma_Q^{(1)}&=&Y_t+Y_b-\frac{8}{3}\alpha_3-\frac{3}{2}\alpha_2-
\frac{1}{30}\alpha_1, \\
\gamma_U^{(1)}&=&2Y_t-\frac{8}{3}\alpha_3-\frac{8}{15}\alpha_1, \\
\gamma_D^{(1)}&=&2Y_b-\frac{8}{3}\alpha_3-\frac{2}{15}\alpha_1, \\
\gamma_L^{(1)}&=&Y_\tau -\frac{3}{2}\alpha_2-\frac{3}{10}\alpha_1, \\
\gamma_E^{(1)}&=&2Y_\tau -\frac{6}{5}\alpha_1, \\
\gamma_{H_1}^{(1)}&=&3Y_b+Y_\tau -\frac{3}{2}\alpha_2-\frac{3}{10}\alpha_1,
 \\ \gamma_{H_2}^{(1)}&=&3Y_t -\frac{3}{2}\alpha_2-\frac{3}{10}\alpha_1.
\end{eqnarray}
Consequently, the renormalization group $\beta $ functions are
\begin{eqnarray}
\beta_{\alpha_i}^{(1)}&=&b_i\alpha_i^2, \\
\beta_{Y_t}^{(1)}&=&Y_t(6Y_t+Y_b-\frac{16}{3}\alpha_3-3\alpha_2
-\frac{13}{15}\alpha_1),\\
\beta_{Y_b}^{(1)}&=&Y_b(Y_t+6Y_b+Y_\tau -\frac{16}{3}\alpha_3-3\alpha_2
-\frac{7}{15}\alpha_1),\\
\beta_{Y_\tau }^{(1)} &=&Y_\tau (3Y_b+4Y_\tau -3\alpha_2
-\frac{9}{5}\alpha_1),\\
\beta_{\mu^2}^{(1)}&=&\mu^2(3Y_t+3Y_b+Y_\tau -3\alpha_2-\frac{3}{5}\alpha_1).
\end{eqnarray}
This allows us immediately to write down the soft term renormalizations
\begin{eqnarray}
\beta_{A_t}^{(1)}&=&6Y_tA_t+Y_bA_b+\frac{16}{3}\alpha_3m_{A_3}+
3\alpha_2m_{A_2} +\frac{13}{15}\alpha_1m_{A_1},\\
\beta_{A_b}^{(1)}&=&Y_tA_t+6Y_bA_b+Y_\tau A_\tau
+\frac{16}{3}\alpha_3m_{A_3}+3\alpha_2m_{A_2}+\frac{7}{15}\alpha_1m_{A_1},\\
\beta_{A_\tau }^{(1)} &=&3Y_bA_b+4Y_\tau A_\tau
+3\alpha_2m_{A_2}+\frac{9}{5}\alpha_1m_{A_1},\\
\beta_{B}^{(1)} &=&3Y_tA_t+3Y_bA_b+Y_\tau A_\tau
+3\alpha_2m_{A_2}+\frac{3}{5}\alpha_1m_{A_2}, \\
\beta_{m_{A_i}}^{(1)}&=&\alpha_ib_im_{A_i}, \\
\beta_{m_{Q}^2}^{(1)}&=&Y_t(m_Q^2+m_U^2+m_{H_2}^2+A_t^2)+
Y_b(m_Q^2+m_D^2+m_{H_1}^2+A_b^2)\\
\nonumber && -\frac{16}{3}\alpha_3m_{A_3}^2-
3\alpha_2m_{A_2}^2-\frac{1}{15}\alpha_1m_{A_1}^2, \\
\beta_{m_{U}^2}^{(1)}&=&2Y_t(m_Q^2+m_U^2+m_{H_2}^2+A_t^2)
-\frac{16}{3}\alpha_3m_{A_3}^2-\frac{16}{15}\alpha_1m_{A_1}^2, \\
\beta_{m_{D}^2}^{(1)}&=&2Y_b(m_Q^2+m_D^2+m_{H_1}^2+A_b^2)
-\frac{16}{3}\alpha_3m_{A_3}^2-\frac{4}{15}\alpha_1m_{A_1}^2, \\
\beta_{m_{L}^2}^{(1)}&=&Y_\tau (m_L^2+m_E^2+m_{H_1}^2+A_\tau^2)-
3\alpha_2m_{A_2}^2-\frac{3}{5}\alpha_1m_{A_1}^2, \\
\beta_{m_{E}^2}^{(1)}&=&2Y_\tau (m_L^2+m_E^2+m_{H_1}^2+A_\tau^2)
-\frac{12}{5}\alpha_1m_{A_1}^2,\\
\beta_{m_{H_1}^2}^{(1)}&=&3Y_b(m_Q^2+m_D^2+m_{H_1}^2+A_b^2)+
Y_\tau (m_L^2+m_E^2+m_{H_1}^2+A_\tau^2)\nonumber \\ &&-
3\alpha_2m_{A_2}^2-\frac{3}{5}\alpha_1m_{A_1}^2, \\
\beta_{m_{H_2}^2}^{(1)}&=&3Y_t(m_Q^2+m_U^2+m_{H_2}^2+A_t^2)-
3\alpha_2m_{A_2}^2-\frac{3}{5}\alpha_1m_{A_1}^2,
\end{eqnarray}
which perfectly coincide with those of ref.(\cite{Boer})

The RG $\beta $ functions for the gauge couplings in the MSSM
are
\begin{eqnarray}
\beta_{\alpha_i}&=&b_i\alpha_i^2+\alpha_i^2\left(\sum_jb_{ij}\alpha_j-\sum_f
a_{if}Y_f \right) \\
&& \hspace{-1cm} +\alpha_i^2\left[\sum_{jk}b_{ijk}\alpha_j\alpha_k
-\sum_{jf}a_{ijf}\alpha_jY_f+\sum_{fg}a_{ifg}Y_fY_g \right] + ... ,
\nonumber
\end{eqnarray}
where $Y_f$ means $Y_t,Y_b$ and $Y_\tau $ and the coefficients
$b_i,b_{ij},a_{if},b_{ijk},a_{ijf}$ and $a_{ifg}$ are given in
ref.~\cite{Ferreira}.

For the gaugino masses we have
\begin{eqnarray}
\beta_{m_{A_i}}&=&b_i\alpha_im_{A_i}+
\alpha_i\left(\sum_jb_{ij}\alpha_j(m_{A_i}+m_{A_j})-\sum_f
a_{if}Y_f(m_{A_i}-A_f) \right) \nonumber \\
&&+\alpha_i\left[\sum_{jk}b_{ijk}\alpha_j\alpha_k(m_{A_i}+m_{A_j}+m_{A_k})
-\sum_{jf}a_{ijf}\alpha_jY_f(m_{A_i}+m_{A_j}-A_f)\right. \nonumber \\
&& \left. +\sum_{fg}a_{ifg}Y_fY_g (m_{A_i}-A_f-A_g) \right] + ... .
\end{eqnarray}

\section{Finiteness of Soft Parameters in a Finite SUSY GUT}

Consider now the application of the proposed formulae to construct
totally finite softly broken theories.

For rigid N=1 SUSY theories there exists a general method of
constructing totally all loop finite gauge theories  proposed in
refs.~\cite{EKT,Jones1,Sibold}.  The key issue of the method is the
one-loop finiteness. If the theory is one-loop finite and satisfies
some criterion verified in one loop~\cite{Kazakov}, it can be
made finite in any loop order by fine-tuning of the Yukawa couplings
order by order in PT. In case of a simple gauge group the Yukawa
couplings have to be chosen in the form
\begin{equation}
Y_a(\alpha)=c^a_1\alpha +c^a_2\alpha^2+..., \label{yuk}
\end{equation}
where the finite coefficients  $c^a_n$ are calculated algebraically in
the n-th order of perturbation theory.

Suppose now that a rigid theory is made finite to all orders by the
choice of the Yukawa couplings as in eq.(\ref{yuk}).
This means that all the anomalous dimensions and the $\beta $ functions
on the curve $Y_a=Y_a(\alpha)$ are identically equal to zero.

Consider the renormalization of the soft terms.  According to
eqs.(\ref{ma}-\ref{m2}) and (\ref{d1},\ref{d2}) the renormalizations of
the soft terms are not independent but are given by the differential
operators acting on the same anomalous dimensions.  One has either
\begin{equation}
\beta_{soft} \sim D_1\gamma (Y,\alpha ), \ \ \ \ {\rm or}
 \ \ \ \beta_{soft} \sim D_2\gamma (Y,\alpha ), \label{b1}
\end{equation}
where  $ \gamma (Y_a,\alpha ) $ is some anomalous dimension.

From the requirement of finiteness
\begin{equation}
\gamma_i(Y_a(\alpha),\alpha )=0,
\end{equation}
the Yukawa couplings $Y_a(\alpha )$ are found in the form (\ref{yuk}).

To reach the finiteness of all the soft terms in all
loop orders, one has to choose the soft parameters
$A_a$ and $m^2_i$ in a proper way. The following statement is
valid:~\cite{Finlet}

\begin{statement}
{\it The soft term $\beta $ functions become equal to zero
if the parameters $A_a$ and $m^2_i$ are chosen in the following form:
\begin{eqnarray} A_a(\alpha )&=&-m_A\alpha
\frac{\partial }{\partial \alpha }\ln Y_a(\alpha ), \label{fina}\\
m^2_i&=&-m_AK^{-1}_{ia}\frac{\partial }{\partial \alpha }\alpha A_a(\alpha )
\label{finm}          \\
&=&m_A^2K^{-1}_{ia}\frac{\partial }{\partial \alpha }\alpha^2
\frac{\partial }{\partial \alpha }\ln Y_a(\alpha ), \nonumber
\end{eqnarray}
where the matrix $K^{-1}_{ia}$ is the inverse of the matrix $K_{ai}$.}
\end{statement}

This statement follows from the form of the operators $D_1$ and $D_2$. After
substitution of solutions (\ref{fina}) and (\ref{finm}) into $D_1$
and $D_2$, $D_1$ becomes a total derivative over $\alpha $ and $D_2$
becomes a second total derivative.

Indeed, consider eq.(\ref{b1}). For $A_a$ chosen as in eq.(\ref{fina}) the
differential operator $D_1$ takes the form
\begin{eqnarray*}
D_1=m_A\alpha \frac{\partial }{\partial \alpha }-
A_aY_a\frac{\partial }{\partial Y_a}=
m_A(\frac{\partial \ln  Y_a}{\partial \ln \alpha }
\frac{\partial }{\partial \ln Y_a} +
\frac{\partial }{\partial \ln \alpha })=m_A\frac{d}{d\ln \alpha }.
\end{eqnarray*}
Hence, since on the curve $Y_a=Y_a(\alpha )$ the anomalous dimension
$\gamma (Y_a,\alpha )$ identically vanishes, so does its derivative
$$\frac{d}{d\ln \alpha }\gamma (Y_a(\alpha ),\alpha ) =0.$$
The  operator $D_2$  is the second derivative.  Using eq.(\ref{finm})
one has
\begin{eqnarray*}
D_2&=&(m_A\alpha \frac{\partial }{\partial \alpha }-
A_aY_a\frac{\partial }{\partial Y_a})^2+ m_A^2\alpha \frac{\partial
}{\partial \alpha }+ m_i^2K_{ai}Y_a\frac{\partial }{\partial Y_a}\\
&=& (m_A\alpha \frac{\partial }{\partial \alpha }- A_aY_a\frac{\partial
}{\partial Y_a})^2 - m_A\frac{\partial A_a}{\partial \ln \alpha }
Y_a\frac{\partial }{\partial Y_a} +m_A( m_A\alpha \frac{\partial
}{\partial \alpha }- A_aY_a\frac{\partial }{\partial Y_a})\\
& =& m_A^2\frac{d}{d \ln \alpha }+m_A^2\frac{d^2}{d\ln^2 \alpha}.
\end{eqnarray*}
The term with the derivative of $A_a$ is essential to
get the total second derivative over $\alpha $, since in the bracket
the derivative $\alpha  \partial/\partial \alpha $  does not act on $A_a$
by construction.
Like in the previous case the total derivatives identically
vanish on the curve $Y_a=Y_a(\alpha )$.

The solutions (\ref{fina},\ref{finm}) can be checked perturbatively
order by order.
In the leading order one has
\begin{equation}
A_a=-m_A,  \ \ \ m^2_i=\frac{1}{3}m_A^2,   \label{oneloop}
\end{equation}
since $\sum_aK^{-1}_{ia}=1/3$. These relations coincide with the already
known ones~\cite{Mezinchescu} and with those coming from
supergravity~\cite{Nilles} and supersting-inspired
models~\cite{Ibanez}. There  they  usually
follow from the requirement of finiteness of the cosmological
constant and probably have the same origin. Note that since the
one-loop finiteness of a rigid theory automatically leads to the
two-loop one and hence the coefficients $c^a_2=0$, the same statement
is valid due to eqs.(\ref{fina},\ref{finm}) for a softly broken theory.
Namely, relations (\ref{oneloop}) are valid up to two-loop order
in accordance with ~\cite{Jones}.
In higher orders, however, they have to be modified.

This way one can construct all loop finite
N=1 SUSY GUT including the soft SUSY breaking terms with the
fine-tuning  given by exactly the same functions as in a rigid theory.

\section{Discussion}

Our approach is based on a consideration of the soft theory as a rigid
one embedded into the external $x-$independent superfields, that are
the charges and masses of the theory. The Feynman rules together with
the operator constructions $D_1$ and $D_2$ are just the
technical consequences of this approach.  One can see that all
the essential information concerning the renormalization is actually
contained in a rigid theory.  There is no independent renormalization
in softly broken SUSY theories.

There might be several applications of this approach. First, one can
use it for finding of the soft terms RG functions in the MSSM and
other theories. Second, one cab try to consider the extended SUSY
theories, like N=2 SUSY, where in unbroken case the exact results are
known. Then one can break N=2 to N=1 or to N=0 in a soft way and
consider the broken theory  along the lines advocated above preserving
the symmetry properties of the exact solution. In fact presumably any
spontaneously broken theory may be treated this way.

Later there has been considerable activity   exploring the
renormalization group invariant relations between the soft term
renormalizations ~\cite{Jones3,Jones4,Zoupanos}. They essentially use
the explicit form of the differential operators $D_1$ and $D_2$
and their reduction to the total derivatives found in~\cite{Finlet}.
In particular the exact form for the correction due to the
$\epsilon$-scalar mass in component approach in the renormalization
scheme corresponding to the NSVZ $\beta $ function has been
obtained~\cite{Jones4}.    The exact $\beta $ function for the scalar
masses in NSVZ scheme for softly broken SUSY QCD has been also
found~\cite{Zoupanos}.


\begin{thebibliography}{99}
\bibitem{supergraph} R. Delbourgo, Nuovo Cim. {\bf 25A} (1975) 646;\\
A. Salam and J. Strathdee, Nucl. Phys. {\bf B86} (1975) 142; \\
K. Fujikawa and W. Lang, ibid. Nucl. Phys. {\bf B88} (1975) 61.
\bibitem{supergraph2} M.T. Grisaru, M. Ro\v{c}ek and W. Siegel, Nucl.
Phys. {\bf B59} (1979) 429.
\bibitem{spurion} L.  Girardello and  M.T. Grisaru,
Nucl. Phys. {\bf B194} (1982) 65.
\bibitem{spurion2} J.A. Helay\"el-Neto, Phys. Lett. {\bf 135B} (1984)
78; \\
F. Feruglio, J.A. Helay\"el-Neto and F. Legovini, Nucl. Phys. {\bf
B249} (1985) 533;
\bibitem{Scholl} M. Scholl, Z. Phys. {\bf C28} (1985) 545.
\bibitem{Yamada} Y. Yamada, Phys. Rev. {\bf D50} (1994) 3537.
\bibitem{AKK} L.A.Avdeev, D.I.Kazakov and I.N.Kondrashuk,
Nucl.Phys. {\bf B510} (1998) 289 (hep-ph/9709397).
\bibitem{Mezinchescu} D.R.T.Jones, L.Mezinchescu and Y.-P. Yao,
Phys.Lett., {\bf 148B} (1984) 317.
\bibitem{Jones} I. Jack and D.R.T. Jones, Phys. Lett. {\bf 333B}
(1994) 372.
\bibitem{Finlet} D.I.Kazakov, Phys.Lett. {\bf B421} (1998) 211
(hep-ph/9709465)
\bibitem{Shifman} J. Hisano and M.A. Shifman, Phys.Rev.
{\bf D56} (1997) 5475 (hep-ph/9705417).
\bibitem{NewJ} I. Jack and D.R.T. Jones, Phys.Lett. {\bf B415}
(1997) 383 (hep-ph/9709364).
\bibitem{Boer}  see e.g.
V. Barger, M.S. Berger and P. Ohmann, Phys. Rev. {\bf D47} (1993) 1093;
\\ W. Boer, R. Ehret and D.I. Kazakov, Z. Phys. {\bf C67} (1995) 667;
\bibitem{Ferreira}  P.M. Ferreira, I. Jack and D.R.T. Jones,
                  hep-ph/9605440, Phys. Lett. {\bf 387B} (1996) 80.
\bibitem{EKT}  A.V.Ermushev, D.I.Kazakov and O.V.Tarasov, Nucl.Phys.,
{\bf B281} (1987) 72.
\bibitem{Jones1}  D.R.T.Jones, Nucl.Phys. {\bf B277} (1986) 153.
\bibitem{Sibold} C.Lucchesi, O.Piguet and K.Sibold, Phys.Lett. {\bf
201B} (1988) 241.
\bibitem{Kazakov} D.I.Kazakov, Mod.Phys.Lett., {\bf A9} (1987) 663.
\bibitem{Nilles} H.-P.Nilles, Phys.Rep. {\bf 110} (1984) 1.
\bibitem{Ibanez} A.Brignole, L.E.Ib\'a\~nez and C.Mu\~noz, Nucl.Phys.
{\bf B422} (1994) 125.
\bibitem{Jones3} I.Jack, D.R.T.Jones,  A.Pickering, hep-ph/9712542.
\bibitem{Jones4} I.Jack, D.R.T.Jones,  A.Pickering,  hep-ph/9803405.
\bibitem{Zoupanos} T.Kobayashi, J.Kubo and G.Zoupanos, hep\--ph/9802267.
\end{thebibliography}
\end{document}